\documentstyle[preprint,aps]{revtex}
\begin{document}
\preprint{SNUTP-94-47}
\draft
\title{ Exact  Propagator  for  $SU(N)$ Coherent State}
\author{Phillial Oh\cite{poh}}
\address{Department of Physics,
Sung Kyun Kwan University,
Suwon 440-746,  KOREA}

\maketitle

\begin{abstract}
We present  a classical integrable  model of $SU(N)$ isospin
defined on complex projective phase space in the
 external  magnetic field and solve it exactly by
constructing the action-angle variables for the system.
We quantize the system using
the coherent state path integral method and obtain an exact
expression for quantum mechanical propagator  by
solving the time-dependent Schr\"odinger equations.
\end{abstract}


There exist a number of integrable systems\cite{arno,abra}
 which can be solved exactly.
Classical solvability is a question of finding the action-angle
variables of the system.
Quantum mechanically, one way of solving the system would be to
find an exact expression for the propagator by solving the
time-dependent Schr\"odinger equation.

The classical models of point particles carrying isospin charge
in interaction with external gauge field have been investigated
for over two decades\cite{wong}.
 Recently, attetion has been paid to them  in relation to the
 non-Abelian Chern-Simons quantum mechanics
\cite{verl,lo,lo2,bak1,kimoh} which may have some application
 in condensed matter physics\cite{ston}.
In this Letter, we construct  a  completely integrable
classical model of non-relativistic
$SU(N+1)(N\geq 1)$ isospin particle
interacting with  constant  external magnetic field and
show that it can be solved exactly in the above way.
We will be concerned with only isospin degrees of freedom
and drop the spatial dependence all togather.

Classical $SU(N+1)$ symmetry can be well described on
  complex projective space
${\cal M}=CP(N)$ which is a
symplectic manifold and therefore could be considered to be the
phase space of classical mechanics. The symplectic structure
is given by the Fubini-Study metric \cite{grif} with
a  constant $J$
\begin{equation}
\Omega=2iJ\left[\frac{d{\bar\xi}\wedge d\xi}{1+
\vert\xi\vert^2}-\frac{( \xi d{\bar\xi})\wedge({\bar\xi}d\xi)}
{(1+\vert\xi\vert^2)^2}\right],
\label{omeg}
\end{equation}
where $\xi=(\xi_1,\cdots,\xi_N)$ is the coordinates of $CP(N)$
and $\vert\xi\vert^2=\sum_{m=1}^N\xi^*_m\xi_m\equiv\bar\xi\xi.$
We introduce the notation
$\xi^A=(\xi^*_m, \xi_n)$ $(m,n=1,\cdots,N)$
and write the symplectic two form Eq.(\ref{omeg})
as $\Omega=\frac{1}{2}\Omega_{AB}d\xi^A\wedge
d\xi^B$. Then, the Poisson bracket between $ F,H\in
C^\infty({\cal M})$ is defined by
\begin{equation}
\{F,H\}=\Omega^{AB}\partial_AF\partial_BH \label{poi}
\end{equation}
where $\Omega^{AB}$ is the inverse
matrix of  $\Omega_{AB}$.
The use of Eq.(\ref{omeg}) results in the following
expression for Poisson bracket:
\begin{equation}
\{F,H\}=-i\sum_{m,n}g^{mn}\left(\frac{\partial F}
{\partial \xi^*_m}\frac{\partial H}{\partial \xi_n}-
\frac{\partial F}{\partial\xi_n}\frac{\partial H}
{\partial  \xi^*_m}\right),
\end{equation}
where $g^{mn}$ is the inverse of the Fubini-Study metric
given by
\begin{equation}
g^{mn}=\frac{1}{2J}(1+\vert\xi\vert^2)
(\delta_{mn}+ \xi^*_m\xi_n).\label{metric}
\end{equation}

Now, the group $SU(N+1)$ acts transitively on $ CP(N)$
and leaves the symplectic two form Eq.(\ref{omeg}) invariant.
It can be shown\cite{kimoh}  that
the group action $g=\exp(it T^a)\in SU(N+1), t\in {\bf R}$
generates the following Hamiltonian vector field on ${\cal M}$
\begin{equation}
X_a=i\left[(T^a)_{mo}+(T^a)_{mn}\xi_n-(T^a)_{00}\xi_m-(T^a)_{0n}
\xi_n\xi_m\right]\frac{\partial}{\partial \xi_m} + (c.c).
\label{vector}
\end{equation}
Here $T^a$'s form the generators of the Lie algebra ${\cal G}=
su(N+1)$:
\begin{equation}
[T^a, T^b]=if^{abc}T^c
\end{equation}
We use the normalization $Tr(T^aT^b)=(1/2)\delta_{ab}$ and
 the notation that $T^a$ is $(N+1)\times (N+1)$
matrix with element $(T^a)_{ij}$ $(i,j=0,1,\cdots,N)$.
The above Hamiltonian vector field defines the isospin function
(so-called momentum mapping
function\cite{abra}) $Q^a$ on ${\cal M}$ by
\begin{equation}
X_a\rfloor \Omega+dQ^a=0.\label{moment}
\end{equation}
One can show that $Q^a$ can be
 expressed as follows\cite{lo2,kimoh}
\begin{equation}
Q^a=-2J\sum_{i,j=0}^{N} u^*_i(T^a)_{ij}u_j{\Big\vert}
_{u_0=\frac{1}
{\sqrt{1+\mid\xi\mid^2}}, u_m=u_0\xi_m}.\label{def3}
\end{equation}
It  can be easily  checked
that $X_a^\prime$s form  $su(N+1)$ Lie algebra $[X_a,X_b]=
f^{abc}X_c$ and $Q^a$ satisfy
$su(N+1)$ Poisson-Lie algebra :
\begin{equation}
\{Q^a,Q^b\}=f^{abc}Q^c.\label{op}
\end{equation}

To define  our integrable model with the use of isospin functions
Eq.(\ref{def3}) and (\ref{op}), we need an explicit representation
for the generator $T^{a\prime}$s.
Recall that  $T^{a\prime}$s can be decomposed into a maximally
commuting set $\{H^m\} (m=1,\cdots, N)$ which is called the Cartan
subalgebra and a set of ladder operators $E_\alpha$ one for each
root vectors $\alpha$.
We use the defining representation of $SU(N+1)$ where the diagonal
$H^{m\prime}$s are
defined as follows\cite{geor}:  $H^m$ has $m$ $1^\prime$s along
the  diagonal from the upper
left-hand corner. The next diagonal element is $-m$ to make it
traceless. The rest of the diagonal element (if any) are zero.
So with the normalization defined previously, we have
\begin{equation}
(H^m)_{ij}= (\sum_{k=0}^{m-1}\delta_{ik}\delta_{jk}-
m\delta_{i,m}\delta_{j,m})/\sqrt{2m(m+1)}
\end{equation}
The vector field generated by  $\exp(itH^m)$ can be expressed
using Eq.(\ref{vector}) as
\begin{equation}
X_m=\frac{-i}{\sqrt{2m(m+1)}}\left[
(m+1)\xi_m\frac{\partial}{\partial\xi_m}
+\left(\sum_{k=m+1}^N\xi_k\frac{\partial}{\partial\xi_k}\right)
\right]+(c.c)
\end{equation}
In the above, the index $m$ is not summed.
The corresponding isospin functions Eq.(\ref{def3}) are given by
\begin{equation}
Q^m  =\frac{-2J}{\sqrt{2m(m+1)}}(\sum_{k=0}^{m-1} u^*_ku_k
-m u^*_mu_m)
{\Big\vert}_{u_0=\frac{1}{\sqrt{1+\vert\xi\vert^2}}, u_m=u_0\xi_m}.
\label{def2}
\end{equation}

We define an integrable model on ${\cal M}$ with the Hamiltonian
given by
a linear combination of the above  isospin functions
\begin{equation}
H  = \sum_{m=1}^N \mu_mQ^m.\label{hamil}
\end{equation}
with real constants $\mu_m$.
When restricted to $SU(2)$ case with $N=1$, this model describes
a classical (iso)spin  in the constant external magnetic field as
 can be readily checked in terms of stereo graphical projection
$\xi=\tan (\theta/2)e^{-i\phi}$:
\begin{equation}
H=\mu  Q, \quad Q =-J\cos\theta.
\end{equation}
with the magnetic field given by $\mu$ and $J$ represents
the magnitude of classical (iso)spin.

By construction, there are $N$ conserved quantities $Q^m$ which
are in involution
\begin{equation}
\{H, Q^m\}=0,\quad \{Q^m, Q^n\}=0
\end{equation}
and the system is completely integrable\cite{arno,abra}.
The Liouville's theorem states that the manifold ${\cal M}_Q$
defined by the level set of the functions $Q^m$,
\begin{equation}
{\cal M}_Q=\{x:Q^m(x)=q^m, x=(\bar\xi,\xi)\}
\end{equation}
with $q^m$=constant,
is a smooth $N$-dimensional manifold  diffeomorphic to the torus
\begin{equation}
T^N=\{(\phi_1, \cdots, \phi_N)mod \ \ 2\pi\}.
\end{equation}
and invariant with respect to the phase flow with the Hamiltonian.
In this case, we can find the action variables $(I_1,\cdots,I_N)$
conjugate to the angle $(\phi_1, \cdots, \phi_N)$
so that the original symplectic structure
Eq.(\ref{omeg}) is expressed by the canonical two form
\begin{equation}
\Omega=\sum_{m=1}^NdI_m\wedge d\phi_m.\label{cano}
\end{equation}

The explicit form of the action variables can be found by the use
of stereo graphical projection. Let us introduce the polar angles
$(\theta_1,\cdots,\theta_N) (0\leq\theta\leq \pi)
$ and $(\phi_1, \cdots, \phi_N)$ via
\begin{eqnarray}
\xi_1 & = & \tan (\theta_1/2)\cos(\theta_2/2)e^{-i\phi_1}
\nonumber\\
\xi_2 & = & \tan (\theta_1/2)\sin(\theta_2/2)\cos(\theta_3/2)
e^{-i\phi_2}\nonumber\\
     & \cdot & \cdots\nonumber\\
\xi_{N-1} & = & \tan (\theta_1/2)\sin(\theta_2/2)\cdots
\sin(\theta_{N-1}/2)\cos(\theta_N/2)e^{-i\phi_{N-1}}\nonumber\\
\xi_{N} & = & \tan (\theta_1/2)\sin(\theta_2/2)\cdots
\sin(\theta_{N-1}/2)\sin(\theta_N/2)e^{-i\phi_{N}}\label{polar}
\end{eqnarray}
It is interesting to find that the action variables  $I_m$
are given by the following formula:
\begin{eqnarray}
I_m & = &2J\sin^2(\theta_1/2)  \cdots \sin^2(\theta_m/2)
\cos^2(\theta_{m+1}/2) \quad (m<N)\nonumber\\
I_N & = & 2J\sin^2(\theta_1/2)\cdots
\sin^2(\theta_{N-1}/2)\sin^2(\theta_N/2)
\end{eqnarray}
It can be expressed more compactly as $I_m=2J\xi^*_m\xi_m
/(1+\vert\xi\vert^2)$.
One can  check  that substitution of Eq.(\ref{polar}) into
Eq.(\ref{omeg}) produces Eq.(\ref{cano}).
The above action variables enables $Q^m$ to be
expressed as a linear combination of them:
\begin{equation}
Q^m=\frac{1}{\sqrt{2m(m+1)}}\left[(m+1)I_m+\sum_{k=m+1}^N
I_k-2J\right]
\end{equation}
Substituting into Eq.(\ref{hamil}), we obtain
\begin{equation}
H=\sum_{n=1}^N\omega_nI_n +H_0\label{ham}
\end{equation}
where $H_0$ is a constant and the Larmor frequency
$\omega_n$ is given by
\begin{equation}
\omega_n=\sum_{m=1}^N\frac{\mu_m}{\sqrt{2m(m+1)}}
\left[(m+1)\delta_{nm}+\theta_{nm}\right]
\label{larmor}
\end{equation}
with the step symbol $\theta_{nm}$,
\begin{equation}
\theta_{nm}=\left\{ \begin{array}{ll}
	  0,  &  n\leq m \\
	   1,  &  n> m.
	       \end{array} \right.
\end{equation}
With Eq.(\ref{ham}), we have completely solved the system
and the solution is given by
\begin{equation}
\theta_m(t)=\theta_m(0), \quad \phi_m(t)=\phi_m(0)+\omega_mt.
\label{sol}
\end{equation}

Instead of performing the path integral in terms of the above
Darboux variables\cite{niel,alek2},  we quantize
the above classical model using the coherent state
path integral method\cite{klau1,pere1,zhan}
and  obtain the exact quantum mechanical propagator
by solving the time-dependent Schr\"odinger equations.
We first construct coherent states on $CP(N)$. Let us consider
$\vert 0\rangle$, the highest weight state annihilated by all
positive roots of $SU(N+1)$ algebra in Cartan basis.
Then for $CP(N)$ with given $P
\equiv 2J$ $(P\in {\bf Z}^+)$ we have an irreducible
representation  $(P,0,\cdots,0)$ of $SU(N+1)$ group\cite{kiri}
and there are precisely  $N$ negative roots
$E_\alpha, \alpha=1,2,\cdots,N$
 such that $E_\alpha\vert 0\rangle\neq\vert 0\rangle.$
 Let us label $\{E_\alpha\}=\{E_m\}$.
We define a coherent state on $CP(N)$ corresponding
to the point $\xi=(\xi_1,\cdots,\xi_N)$ by \cite{pere1,zhan}
\begin{equation}
\vert P,\xi\rangle=\exp(\sum_m\xi_mE_m)\vert 0\rangle
\end{equation}
Notice that this definition differs from the usual one by
the normalization factor.
We have chosen this definition here because in the subsequent
analysis, $\bar\xi$ and $\xi$
can be treated independently and the overspecification problem
can be side-stepped\cite{klau,fadd2,brow}. We denote
$\vert P,\xi\rangle=\vert\xi\rangle$ from now on.
The coherent states which we have defined on
 $CP(N)$ have the
following two  properties which are essential
in the path integral formulation. One is the resolution
of unity,
\begin{equation}
\int D\mu({\bar \xi},\xi)\frac{ \vert
\xi\rangle \langle \xi\vert}{ (1+\vert\xi\vert^2)^{2J}}=I,
\label{id}
\end{equation}
where $D\mu({\bar \xi},\xi)=c d\bar\xi d\xi/
(1+\vert\xi\vert^2)^{N+1}$ with a constant $c$  is the
Liouville measure\cite{zhan}.
The other is reproducing kernel,
\begin{equation}
\langle \xi^{\prime\prime}\vert \xi^\prime\rangle=
(1+\bar\xi^{\prime\prime}\xi^\prime)^{2J}.\label{ker}
\end{equation}

We are interested in evaluating the propagator
\begin{equation}
G(\bar\xi^{\prime\prime},\xi^\prime;t)
= \langle \xi^{\prime\prime}\vert e^{-i{\hat H}t}\vert
\xi^\prime\rangle.
\end{equation}
Inserting Eq.(\ref{id}) and using Eq.(\ref{ker}) repeatedly,
we obtain the following expression
\begin{equation}
G(\bar\xi^{\prime\prime},\xi^\prime;t)=
\int D\mu\exp\left\{2J\log(1+\bar\xi^{\prime\prime}
\xi(t_f))+i\int_{t_i}^{t_f}dt\left[i
\frac{2J{\bar \xi}\dot \xi}{1+\vert\xi\vert^2}
-  H(\bar\xi, \xi)\right]\right\}\label{prop}
\end{equation}
where
$ H(\bar\xi, \xi)=\langle \xi\vert {\hat H}(\bar\xi, \xi)\vert
\xi\rangle/\langle \xi\vert \xi\rangle$ is the classical
Hamiltonian  given by the Eq.(\ref{hamil}).
The boundary conditions in the path integral
is given by $\xi(t_i)=\xi^\prime$ and $\bar\xi(t_f)=
\bar\xi^{\prime\prime}$.
We introduced $ \xi(t_f)$ which is only a  superfluous
variable because the result of path integral
 Eq.(\ref{prop}) does not depend on this variable.
It depends only on
$\bar\xi^{\prime\prime}$ and $\xi^\prime$.

The equations of motion
\begin{equation}
i\dot\xi_m=g^{\ast mn}\frac{\partial H(\bar\xi,\xi)}
{\partial \xi^*_n},\quad  i\dot\xi^*_m=-g^{mn}
\frac{\partial  H(\bar\xi,\xi)}{\partial \xi_n},
\end{equation}
are those of collections of $N$ harmonic oscillator
although the Hamiltonian Eqs.(\ref{hamil}) and (\ref{def2})
appears to be highly nonlinear:
\begin{equation}
 \dot\xi^{\ast}_m-i\omega_m\xi^{\ast}_m=0,
\quad \dot\xi_m+i\omega_m\xi_m=0.\label{equation}
\end{equation}
Here  $\omega_m$ is given by Eq.(\ref{larmor}).
The solutions are given by
\begin{equation}
\xi^{\ast}_m(t)=\xi^{\ast\prime\prime}_me^{i\omega_m
(t-t^{\prime\prime})}\quad
\xi_m(t)=\xi^{\prime}_me^{-i\omega_m(t-t^\prime)}
\label{soll}
\end{equation}
Using the above fact, we can evaluate the propagator
by the semiclassical approximation method \cite{zinn}
and the result agrees\cite{oh} with the exact
propagator which is obtained by solving the time-dependent
Schr\"odinger equation as is adopted in this paper.

In order to have explicit operator form for the Hamiltonian to
set up the Schr\"odinger equation, we resort to geometric
quantization\cite{wood}. In the geometric quantization of
classical phase space ${\cal M}=CP(N)$ with symplectic structure
$\Omega$, we quantize classical observables $F\equiv F(Q^a)$
which are functions of only $Q^{a\prime}$s  satisfying
the Poisson-Lie algebra Eq.(\ref{op}).
The prequantum operator corresponding $Q^a$ is given by
\begin{equation}
\hat Q^a=-iX_a-X_a\rfloor \Theta+Q^a
\end{equation}
where  $\Theta$ is the canonical one form
$\Omega=d\Theta$ given by
\begin{equation}
\Theta=i J\frac{{\bar \xi}d\xi-d{\bar \xi}\xi}
{1+\vert\xi\vert^2}.
\end{equation}
Since our Hamiltonian Eq.(\ref{hamil}) is
linear in $Q^{a\prime}$s,
there is no normal ordering ambiguity. Also  the polarization is
chosen such that $G(\bar\xi^{\prime\prime},\xi^\prime;t)$ is a
function  of $\bar\xi^{\prime\prime}$ but not of
$\xi^{\prime\prime}$. Hence  we get the following
time-dependent Schr\"odinger equation for the propagator:
\begin{equation}
i\frac{\partial}{\partial t}
G(\bar\xi^{\prime\prime},\xi^\prime;t)
= \sum_{m=1}^N
\frac{\mu_m}{\sqrt{2m(m+1)}}\left(\left[
(m+1)\xi^{*\prime\prime}_m\frac{\partial}
{\partial\xi^{*\prime\prime}_m}
+\left(\sum_{k=m+1}^N\xi^{*\prime\prime}_k\frac
{\partial}{\partial\xi^{*\prime\prime}_k}\right)\right]
       -2J\right)
 G(\bar\xi^{\prime\prime},\xi^\prime;t)
\end{equation}
We note that the above equation is invariant under the
gauge transformation of the canonical one form by
$\Theta\rightarrow\Theta+d\Lambda(\bar\xi,\xi)$.
The boundary condition is given by
\begin{equation}
G(\bar\xi^{\prime\prime},\xi^\prime;t){\Big\vert}
_{t\rightarrow 0}=(1+\bar\xi^{\prime\prime}\xi^\prime)^{2J}
\end{equation}
It is remarkable that the solution can be obtained in a closed
simple expression by
\begin{equation}
G(\bar\xi^{\prime\prime},\xi^\prime;t)=\left( 1+
\sum_{m=1}^N\xi^{*\prime\prime}_m\xi^{\prime}_m
\exp\{-i\omega_mt\}\right)
^{2J}\exp\{iJ\sum_{m=1}^N\mu_m/\sqrt{m(m+1)/2}t\}.
\end{equation}
The above results reduces to the exact propagator
and reproduces the Weyl character formula
 for $SU(2)$ case\cite{ston1,alva1,kesk}.

In conclusion, we presented  a classical integrable  model of
$SU(N)$ isospin in interaction with external constant magnetic
field neglecting the spatial dependence completely and solved
it exactly  both classically and quantum mechanically.
Especially, we obtained the exact expression for the
quantum mechanical propagator
by solving the time-dependent Schr\"odinger equation set up
by the method of geometric quantization. This method could be
applied to other integrable models such as
the one in which the Hamiltonian is a
sum of quadratic functions of $Q^m$ of Eq.(\ref{def2}) and
other coadjoint orbits\cite{kiri} of Lie group
including the non-compact
case. These and  related issues will be reported
elsewhere\cite{oh}.

\acknowledgments
This work is supported by the KOSEF through
C.T.P. at S.N.U. and Ministry of Education through the Research
Institute of Basic Science.

\end{document}